\documentclass[10pt,conference]{IEEEtran}
\IEEEoverridecommandlockouts

\usepackage{pgfplots}
\usepackage{prettyref}
\usepackage{tikz}
\usepackage{xcolor,colortbl}
\usepackage{multirow}
\usepackage[ruled, linesnumbered, vlined, commentsnumbered]{algorithm2e}
\usepackage[section]{placeins}
\usepackage{booktabs}
\usepackage{threeparttable}
\usepackage{makecell}
\usepackage{graphicx}
\usepackage[listings,skins,breakable]{tcolorbox}
\usepackage{csquotes}
\usepackage{subcaption}
\usepackage{standalone}
\usepackage[skip=0.5\baselineskip]{caption}
\usepackage{wrapfig}
\usepackage{amsmath}
\usetikzlibrary{pgfplots.statistics}
\usepackage{tcolorbox}
\usepackage{pifont}
\usepackage{balance}
\usepackage[utf8]{inputenc}
\usepackage[T1]{fontenc}
\usepackage{microtype}
\usepackage{amsthm}
\usepackage[normalem]{ulem}
\usepackage{enumitem}
\usepackage{hyperref}
\usepackage{xfp}
\usepackage[ruled, linesnumbered, vlined, commentsnumbered]{algorithm2e}
\newtcolorbox{quotebox}{colback=green!10,boxrule=0.4pt,colframe=black,fonttitle=\bfseries,top=2pt,bottom=2pt}

\mathchardef\mhyphen="2D
\newcommand{\vect}[1]{\boldsymbol{#1}}
\DeclareMathAlphabet\mathbfcal{OMS}{cmsy}{b}{n}

\newbox\aMark
\setbox\aMark\hbox{\begin{pgfpicture}\textcolor{red}{\pgfuseplotmark{o}}\end{pgfpicture}}

\newbox\bMark
\setbox\bMark\hbox{\begin{pgfpicture}\textcolor{red}{\pgfuseplotmark{star}}\end{pgfpicture}}

\newrefformat{fig}{Fig.~\ref{#1}}
\newrefformat{tab}{Table~\ref{#1}}
\newrefformat{sec}{Section~\ref{#1}}
\newrefformat{app}{Appendix~\ref{#1}}
\newrefformat{alg}{Algorithm~\ref{#1}}
\newrefformat{property}{Property~\ref{#1}}
\newrefformat{theorem}{Theorem~\ref{#1}}
\newrefformat{lemma}{Lemma~\ref{#1}}
\newrefformat{corollary}{Corollary~\ref{#1}}
\newrefformat{proposition}{Proposition~\ref{#1}}
\newrefformat{def}{Definition~\ref{#1}}
\newrefformat{eq}{equation~(\ref{#1})}

\definecolor{steel}{rgb}{0, 0.2, 0.9} 

\SetKwInput{kwDeclare}{Declare}

\pgfplotsset{compat=newest}

\pgfplotsset{compat=1.11,
    /pgfplots/ybar legend/.style={
    /pgfplots/legend image code/.code={%
       \draw[##1,/tikz/.cd,yshift=-0.25em]
        (0cm,0cm) rectangle (3pt,0.8em);},
   },
}

\SetCommentSty{mycommfont}

\DeclareMathOperator*{\argmin}{argmin}

\def\signed #1{{\leavevmode\unskip\nobreak\hfil\penalty50\hskip2em
  \hbox{}\nobreak\hfil(#1)%
  \parfillskip=0pt \finalhyphendemerits=0 \endgraf}}

\newsavebox\mybox

  \newcommand{\squart}[4]{\begin{adjustbox}{max width=.1\textwidth}\begin{picture}(100,5)
    {\color{black}\put(0,5){\line(1,0){100}}\color{black}\put(0,5){\line(0,1){10}}\put(50,5){\line(0,1){10}}\put(100,5){\line(0,1){10}}\put(25,5){\line(0,1){5}}\put(75,5){\line(0,1){5}}\put(-2,-8){\LARGE$0$}\put(42,-8){\LARGE$0.5$}\put(96,-8){\LARGE$1$}}\end{picture}\end{adjustbox}}

  \newcommand{\quart}[4]{\begin{adjustbox}{max width=.1\textwidth}\begin{picture}(100,5)
    {\color{black}\put(#1,5){\line(1,0){#2}}\color{black}\put(#1,2){\line(0,1){6}}\color{black}\put(\fpeval{#1+#2},2){\line(0,1){6}}\color{steel}\put(#3,5){\circle*{7}}\color{black}\put(#3,5){\circle{7}}}\end{picture}\end{adjustbox}}

      \newcommand{\quartexp}[4]{\begin{adjustbox}{max width=.1\textwidth}\begin{picture}(20,5)
    {\color{black}\put(#1,3){\line(1,0){#2}}\color{black}\put(#1,0){\line(0,1){6}}\color{black}\put(\fpeval{#1+#2},0){\line(0,1){6}}\color{steel}\put(#3,3){\circle*{4}}\color{black}\put(#3,3){\circle{4}}}\end{picture}\end{adjustbox}}

\def\BibTeX{{\rm B\kern-.05em{\sc i\kern-.025em b}\kern-.08em
    T\kern-.1667em\lower.7ex\hbox{E}\kern-.125emX}}
\begin{document}

\newcommand{\approach}[1]{\texttt{LiDOS} }
\newcommand{\approachd}[1]{\texttt{LiDOS},}
\newcommand{\approachfs}[1]{\texttt{LiDOS}.}
\newcommand{\approachs}[1]{\texttt{LiDOS$_{sta}$} }

\title{Lifelong Dynamic Optimization for Self-Adaptive Systems: Fact or Fiction?}

\author{\IEEEauthorblockN{Tao Chen}
\IEEEauthorblockA{\textit{Department of Computer Science} \\
\textit{Loughborough University, Leicestershire, UK}\\
t.t.chen@lboro.ac.uk}}

\maketitle

\begin{abstract}
When faced with changing environment, highly-configurable software systems need to dynamically search for promising adaptation plan that keeps the best possible performance, e.g., higher throughput or smaller latency --- a typical planning problem for self-adaptive systems (SASs). However, given the rugged and complex search landscape with multiple local optima, such a SAS planning is challenging especially in dynamic environments. In this paper, we propose \approachd, a lifelong dynamic optimization framework for SAS planning. What makes \approach~unique is that to handle the ``dynamic'', we formulate the SAS planning as a multi-modal optimization problem, aiming to preserve the useful information for better dealing with the local optima issue under dynamic environment changes. This differs from existing planners in that the ``dynamic'' is not explicitly handled during the search process in planning. As such, the search and planning in \approach~run continuously over the lifetime of SAS, terminating only when it is taken offline or the search space has been covered under an environment.

Experimental results on three real-world SASs show that the concept of explicitly handling dynamic as part of the search in the SAS planning is effective, as \approach~outperforms its stationary counterpart overall with up to $10\times$ improvement. It also achieves better results in general over state-of-the-art planners and with $1.4\times$ to $10\times$ speedup on generating promising adaptation plans. 

\end{abstract}

\begin{IEEEkeywords}
Self-adaptive systems, search-based software engineering, multi-objectivization, configuration tuning
\end{IEEEkeywords}

\section{Introduction}

Many software systems are highly-configurable, thereby making them flexible to different needs. When operating under dynamic and uncertain environment, those systems are capable of changing their own configuration at runtime with an aim to achieve the best of their performance objective, e.g., higher throughput or smaller latency~\cite{DBLP:conf/splc/LesoilATBJ21,DBLP:journals/csur/ChenBY18} --- a typical type of self-adaptive systems (SASs) that we consider in this work. For example, \textsc{Apache Storm}, a stream processing framework, can change some adaptation options (e.g., \texttt{num\_counters} and \texttt{num\_splitters}) at runtime to react to the changing batch of the jobs with different types and workloads~\cite{nair2018finding,DBLP:conf/mascots/JamshidiC16}.

In SAS, a crucial problem is planning~\cite{DBLP:conf/dagstuhl/LemosGMS10,DBLP:journals/pieee/ChenBY20,DBLP:conf/wosp/0001BWY18,chen2020search},
i.e., what is the next best adaptation plan to take under time-varying environments? From the literature, a promising research direction for tackling this challenges is relying on the paradigm of Search-Based Software Engineering (SBSE), where a search algorithm is used to continuously find the optimal plan for SASs~\cite{Chen2018FEMOSAA,DBLP:conf/sigsoft/ElkhodaryEM10,DBLP:journals/ase/GerasimouCT18,DBLP:conf/icac/RamirezKCM09,DBLP:conf/gecco/0001LY18,DBLP:journals/infsof/ChenLY19,DBLP:journals/taas/KinneerGG21,DBLP:conf/icse/Kumar0BB20}. Indeed, search and optimization is at the fundamental part of the planning, and it is complementary to the other approaches, such as control theoretical~\cite{DBLP:conf/sigsoft/FilieriHM15,DBLP:conf/sigsoft/MaggioPFH17,DBLP:conf/sigsoft/ShevtsovW16} and learning-based~\cite{DBLP:conf/icse/ChenB13,DBLP:conf/icse/HaZ19,DBLP:conf/ucc/ChenBY14,DBLP:journals/tse/ChenB17,DBLP:conf/icse/Chen19b,DBLP:conf/kbse/JamshidiSVKPA17}.

\begin{figure}[t!]
\centering
\includegraphics[width=0.7\columnwidth]{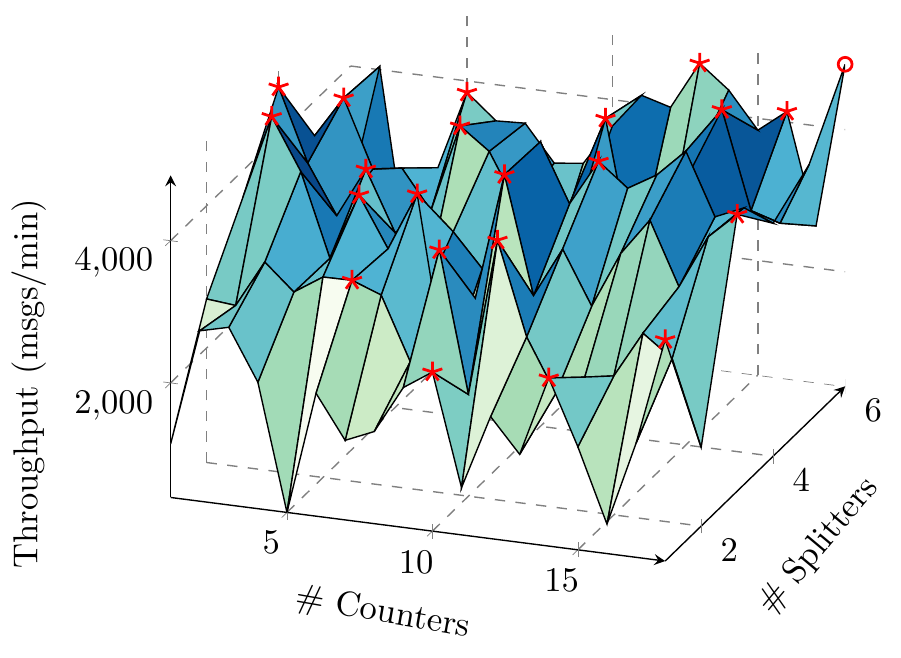}
\caption{A projected landscape of the \textit{Throughput} 
		with respect to adaptation options \texttt{Splitters} and \texttt{Counters} 
		for \textsc{Storm} under the \textsc{RollingCount} environment.	
		\copy\aMark~and \copy\bMark~denote the global and local optimum, respectively.}
\label{fig:exp}
\end{figure}


Yet, apart from the exponentially growing search space and non-linear interaction between the options, what makes the planning particularly difficult is the presence of high sparsity in highly-configurable systems. That is, as shown by Nair \textit{et al.}~\cite{nair2018finding} and Jamshidi \textit{et al.}~\cite{DBLP:conf/mascots/JamshidiC16}, the neighboring adaptation plans can also have drastically different performance given an environment, leading to a highly complex search landscape. For example, we see a rather rugged landscape from Figure~\ref{fig:exp} on the case of \textsc{Storm} under the \textsc{WordCount} environment, where there are many local optima --- the sub-optimal adaptation plans for which all the neighboring plans have worse performance --- that can easily trap the planner. As a result, the produced plans may be far away from the global optimum, especially considering that the planner needs to produce a plan efficiently with little overhead. As we will show in Section~\ref{sec:char}, the landscape can differ following dynamic environmental changes, which further complicates the SAS planning.

In this paper, we tackle the above challenge by proposing \approachd, a framework that implements the concept of lifelong and truly dynamic optimization for SAS planning. By designing a mechanism that takes the characteristics of SAS planning into account, the ``dynamic'' we consider refers to the case that the changes are explicitly captured as the search in the planner proceeds, which is the exact definition of dynamic optimization coined by Nguyen \textit{et al.}~\cite{DBLP:journals/swevo/NguyenYB12}, hence the useful information for planning after the change can be preserved to better overcome local optima. As such, the search and optimization in \approach~runs continuously, terminating only when the SAS is taken offline or all search space has been covered under an environment --- the true lifelong optimization. In this way, we hope to make the lifelong dynamic optimization more factual for SASs. This differs from previous SBSE work for SAS planning where dynamic optimization is more ``fictional'', such that the planner is pseudo-dynamic because, although it runs continuously, the true ``dynamic'' is not explicitly considered in the search, but a vanilla search algorithm is directly adopted and hoping that the changes will eventually be coped with via the re-evaluated fitness~\cite{DBLP:conf/icac/RamirezKCM09,DBLP:conf/gecco/0001LY18,DBLP:journals/infsof/ChenLY19,DBLP:journals/taas/KinneerGG21}; it can also be distinguished from the stationary planners, in which the planning is restarted from scratch upon an environmental change or based on a fixed frequency~\cite{Chen2018FEMOSAA,DBLP:conf/sigsoft/ElkhodaryEM10,DBLP:journals/ase/GerasimouCT18}. Both the pseudo-dynamic and stationary planners are not ideal given the rugged landscape and the need of timely planning for SAS.

The key novelty of this work is that we formulate the lifelong dynamic optimization for SAS as a multi-modal optimization problem. This was inspired by our observation that, for highly-configurable systems, the global optimum of performance in an environment is likely to be (or very close to) a local optimum in the other. Our aim is to continuously preserve as many local optima as possible without losing the tendency towards achieving the global optimum, hence the planning can better deal with the complex landscape even under environmental changes. The core of \approach~for tackling the multi-modal optimization problem is a refined MMO --- the meta multi-objectivization model proposed by Chen and Li at FSE'21~\cite{ChenMMO21}, because (1) it is simple; (2) it fits the needs of a multi-modal optimization problem; and (3) it has been shown to be promising in mitigating local optima traps for highly-configurable systems. In a nutshell, our contributions are:

\begin{enumerate}
    \item We formulate the lifelong dynamic optimization for SAS planning as a multi-modal optimization problem motivated by characteristics of the highly-configurable systems.
    \item To solve the multi-modal optimization problem, we refine the MMO with a new auxiliary objective that can be easily measured, and no normalization is required, thereby eliminating the need of tuning the weight, which has been shown as a highly sensitive parameter in MMO~\cite{ChenMMO21}.
    \item We design an architecture that uses a hierarchical feedback loop to realize the concept of lifelong dynamic optimization for SASs, where the search in planning runs continuously and the dynamic changes that occur during the run are explicitly handled.
    \item Through experiments on three real-world SASs of different performance objectives, scales, complexity, and under six environments, we evaluate \approach~against its stationary variant (for ablation analysis), as well as state-of-the-art pseudo-dynamic~\cite{DBLP:conf/icac/RamirezKCM09,DBLP:journals/taas/KinneerGG21} and stationary planners~\cite{Chen2018FEMOSAA}.
\end{enumerate}

The results are encouraging: we observe that when comparing with its stationary variant, \approach~produces considerably better adaptation plans in general (with $p<0.05$ and high effect size) and up to $10\times$ improvement. \approach~also performs significantly better overall than the state-of-the-art planners with greater efficiency: it shows $1.4\times$ to $5\times$ speedup of planning over the pseudo-dynamic planners and $2.5\times$ to $10\times$ speedup on the stationary one. This proves that lifelong (truly) dynamic optimization for SAS planning is more of a fact than fiction. 

To promote open science, the code and data in this work can be accessed at: \textcolor{blue}{\url{https://doi.org/10.5281/zenodo.5586103}}.

In what follows, 
Section~\ref{sec:prob} introduces the background. 
Section~\ref{sec:method} elaborates the motivation and design of our \approach~framework. 
Section~\ref{sec:study} presents our experiment methodology, 
followed by a discussion of the results in Section~\ref{sec:results}. 
The threats to validity are discussed in Section~\ref{sec:discussion}. 
Sections~\ref{sec:related} and~\ref{sec:con} analyze the related work and conclude the paper, respectively.

\section{Preliminaries}
\label{sec:prob}

Here, we describe the necessary background and context.

\begin{figure}[t!]
\centering
\includegraphics[width=0.8\columnwidth]{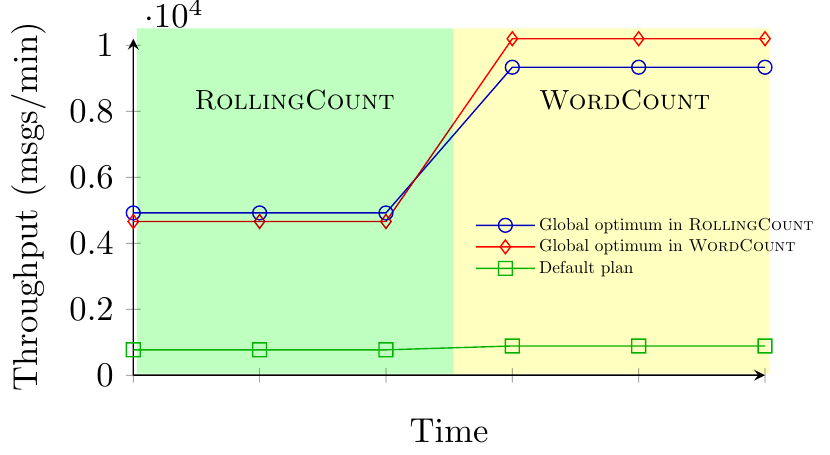}
\caption{Changing throughput on \textsc{Storm} from the \textsc{RollingCount} to \textsc{WordCount} environment (highlighted areas), each of which has different batches of jobs, types and workloads.}
\label{fig:case}
\end{figure}

\subsection{Self-Adaptation for Configurable Systems}


While the purpose of self-adaptation can vary, in this work, we focus on improving the performance of highly-configurable systems such as smaller latency and higher throughput. We model this type of SAS with $n$ adaptation options such that the $i$th option is denoted as $x_i$, which can be either a binary or integer variable.
The search space of all adaptation plans, 
$\mathbfcal{X}$, 
is the Cartesian product of the possible values for all the $x_i$. 
Formally, 
the ultimate goal\footnote{Without loss of generality, we assume minimizing the performance.} is to achieve the following for every monitoring timestep despite the environment changes:
\begin{align}
	\argmin~f(\vect{x}),~~\vect{x} \in \mathbfcal{X}
	\label{Eq:SOP}
\end{align}
where $\vect{x} = (x_1, x_2, ..., x_n)$. However, recall the landscape shown in Figure~\ref{fig:exp}, the planner may be easily trapped at some of the undesired local optima, making the problem non-trivial to address. The need for planning with less overhead (due to large search space and/or expensive measurements) makes it even more challenging. An additional complication with SAS is that, as the environment change, the search landscape can also change, which we will discuss in Section~\ref{sec:char}.

As a concrete example in Figure~\ref{fig:case}, \textsc{Storm} can handle data streaming process under various environments on the incoming batch of jobs, e.g., \textsc{RollingCount} and \textsc{WordCount}. We see that (1) although the default plan leads to different results across the environments, it performs fairly poor overall --- compared with the global optimum, it is $6.4\times$ worse on \textsc{RollingCount} and $11.5\times$ worse on \textsc{WordCount}; (2) the global optimum can differ for different environments. All above motivate the need of self-adaptation for \textsc{Storm}, where the aim is to achieve the best possible throughput by searching the right adaptation plan, e.g., settings for \texttt{num\_counters} and \texttt{num\_splitters}, over changing environments.

\subsection{Dynamic Optimization}

Given the increasingly growing field of SBSE, search algorithms have been widely applied to various software engineering tasks~\cite{Harman2012}. Indeed, any search algorithm is dynamic in nature, i.e., they dynamically determine what direction to explore or which solutions to keep depending on the fitness during the search. However, the active field of dynamic optimization means a different concept. According to Nguyen \textit{et al.}~\cite{DBLP:journals/swevo/NguyenYB12}, dynamic optimization refers to the case where the \textit{``algorithm needs to take into account changes during the optimization process as time goes by.''} The changes here refer to the change in search landscape, including fitness function, global/local optimum and the optimization objectives, etc.

Despite there being a good match between the definition of dynamic optimization and the requirement for SASs, truly dynamic optimization has rarely been explored. Often, the planner with a search algorithm may either be stationary (restart the search whenever upon an environmental change or upon a fixed frequency)~\cite{Chen2018FEMOSAA,DBLP:conf/sigsoft/ElkhodaryEM10,DBLP:journals/ase/GerasimouCT18}; or pseudo-dynamic~\cite{DBLP:conf/icac/RamirezKCM09,DBLP:conf/gecco/0001LY18,DBLP:journals/infsof/ChenLY19,DBLP:journals/taas/KinneerGG21}, i.e., there is no mechanism to handle the potential impact on the search landscape when environmental change occurs, but hoping the nature of the search algorithm can eventually adapt to it using the re-evaluated fitness. Both of the two ways can waste valuable information that would benefit the search after the change, especially considering the highly rugged landscape for SASs from Figure~\ref{fig:exp}. However, the difficulties of truly dynamic optimization for SASs are:
\begin{enumerate}
    \item What information from the planning before the environmental change is useful for planning thereafter?
    \item How to identify and preserve such information?
\end{enumerate}

This paper tackles the above difficulties in the next section.
\section{Designing \approach~}
\label{sec:method}

In this section, we present the motivations and technical designs behind \approachfs.

\subsection{Key Characteristics of SAS Changes}
\label{sec:char}

A clear characteristic for SAS planning from Figure~\ref{fig:exp} is:

\begin{itemize}
    \item \textbf{Characteristic 1:} Highly-configurable systems often come with a multi-modal landscape, i.e., there can be many local optima, making the planner to be easily trapped at them. This is consistent with what has been reported from prior studies on highly-configurable systems~\cite{nair2018finding,DBLP:conf/mascots/JamshidiC16}.
\end{itemize}

\begin{figure}[t!]
\centering
\begin{subfigure}[h]{0.49\columnwidth}
\includegraphics[width=\columnwidth]{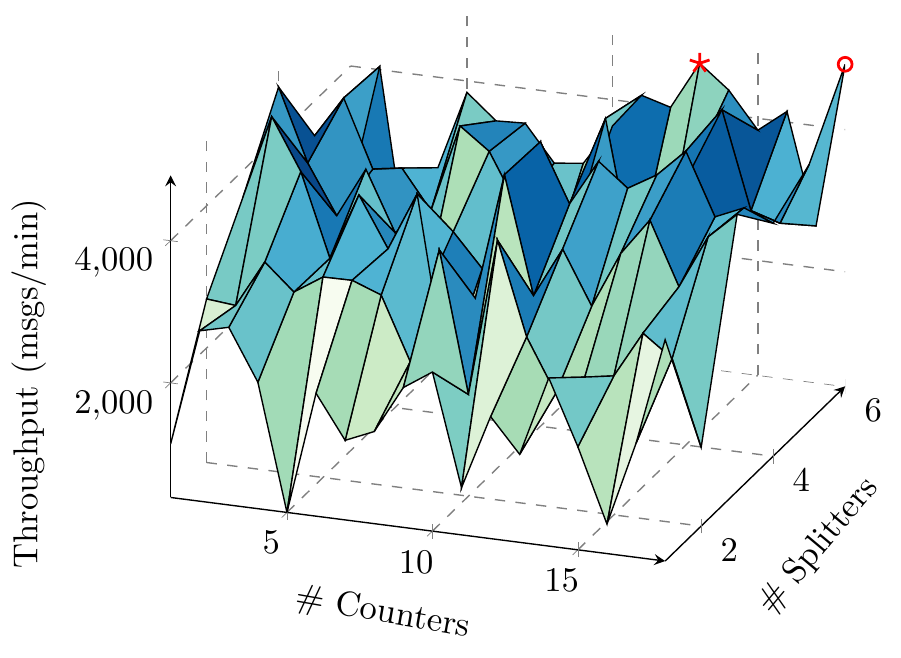}
\subcaption{\footnotesize \textsc{Storm} under \textsc{RollingCount}}
\end{subfigure}
~\hspace{-0.15cm}
\begin{subfigure}[h]{0.49\columnwidth}
\includegraphics[width=\columnwidth]{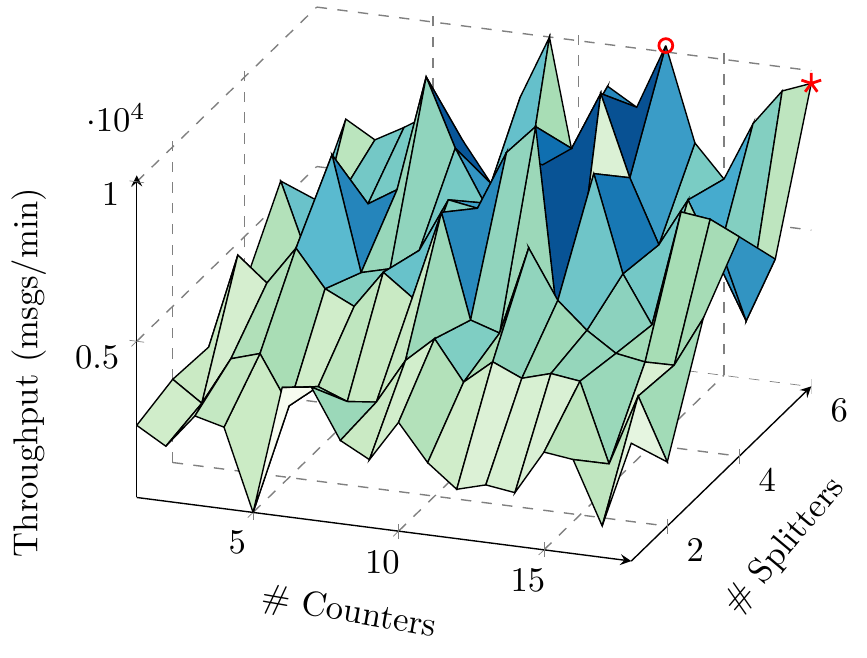}
\vspace{-0.5cm}
\subcaption{\footnotesize \textsc{Storm} under \textsc{WordCount}}
\end{subfigure}
\caption{A projected landscape of the \textit{Throughput} 
		with respect to adaptation options \texttt{Splitters} and \texttt{Counters} 
		for \textsc{Storm} under the \textsc{RollingCount} and \textsc{WordCount} environment.	
		\copy\aMark~denotes the global optimum in an environment and \copy\bMark~denotes a local optimum for the same environment, which is also the global optimum in the other environment.}
\label{fig:exp-land}
\end{figure}

By investigating the performance under different environments of the SAS studied in this work using fitness landscape analysis~\cite{DBLP:conf/gecco/MersmannBTPWR11,DBLP:journals/tsmc/TavaresPC08}, we additionally observed some common patterns and found that (Figure~\ref{fig:exp-land} is an example of \textsc{Storm}):

\begin{itemize}
    \item \textbf{Characteristic 2:} The search landscape does change across different environments, including the global optimum, but the overall multi-modal property is still preserved and there are shared local optima. This matches with the insight by Jamshidi \textit{et al.}~\cite{DBLP:conf/kbse/JamshidiSVKPA17}, 
    which states that the performance values between diverse environments, although different, share similar distribution (an indicator of similar ruggedness/peakedness~\cite{DBLP:conf/gecco/MersmannBTPWR11}), hence the model of the landscape can be linearly transferred.
    
    \item \textbf{Characteristic 3:} The global optimum in one environment is likely to become a local optimum (which is close to the new global optimum) under another environment. This has also been hinted by evidence from the literature~\cite{DBLP:conf/kbse/JamshidiSVKPA17}.

\end{itemize}

The above suggests that the change of environments for SAS can indeed impact the search landscape, but there is certain information on the multi-modality that can be shared before and after the environmental change. This means that truly dynamic optimization that improves the plan continuously is vital for SAS, yet we need a specific mechanism to handle the dynamic changes as the search proceeds and achieve two goals:

\begin{itemize}
    \item \textbf{Goal 1:} preserving as many local optima as possible, since they can provide useful information upon environmental change (\textbf{Characteristic 2} and \textbf{3}).
    \item \textbf{Goal 2:} but doing so without losing the tendency towards finding the global optimum (\textbf{Characteristic 1}).
\end{itemize}

\subsection{Multi-objectivization with Multi-modality}

\textbf{Goal 1} motivates us to formulate the SAS planning as a multi-modal optimization problem~\cite{DBLP:conf/gecco/SinghD06}, for which there exist quite a few search algorithms. However, those algorithms prioritize exploring more local optima while may negatively affecting \textbf{Goal 2}, hence they are ill-suited to our problem. As a result, in \approachd, we refine the meta multi-objectivization (MMO) model proposed by Chen and Li at FSE'21~\cite{ChenMMO21} for achieving \textbf{Goal 1} and \textbf{Goal 2} simultaneously. 

In a nutshell, MMO is a recently proposed way of optimization for configurable and self-adaptive systems. The idea is to ``multi-objectivize'' the problem using an additional auxiliary objective. Unlike existing work that focuses on the algorithm level, MMO works at the more generic level of optimization model which is compatible with different multi-objective search algorithms, e.g., NSGA-II~\cite{Deb2002}. Formally, MMO is defined as\footnote{We use the linear form of MMO, as Chen and Li~\cite{ChenMMO21} have revealed that different forms exhibit little difference in terms of performance.}:
\begin{align}
\begin{split}
\text{minimize}
\begin{cases}
g_1(\vect{x}) = f_t(\vect{x}) + w \times f_a(\vect{x})\\
g_2(\vect{x}) = f_t(\vect{x}) - w \times f_a(\vect{x})\\
\end{cases}
\end{split}
\label{Eq:model}
\end{align}
whereby $g_1(\vect{x})$ and $g_2(\vect{x})$ are the newly transformed objectives to be optimized; $f_t(\vect{x})$ is the target performance objective that is of concerned, e.g., throughput; $f_a(\vect{x})$ is the auxiliary objective of an additional performance attribute such that, under the given context, no stakeholders care about its value, e.g., CPU load. $w$ is a weight parameter that balances the contributions. While MMO was not originally designed for multi-modal problem but for more efficiently mitigating the issues of local optima, the uniqueness lies in that it simultaneously possesses the following properties in the transformed space:

\begin{enumerate}
    \item Because of the Pareto dominance relation on $g_1(\vect{x})$ and $g_2(\vect{x})$, MMO preserves adaptation plans that with very different values of $f_a(\vect{x})$ (despite being similar on $f_t(\vect{x})$) as they tend to be incomparable in the sense of Pareto dominance. This encourages the search to explore a wider range of the search space and hence more likely to find and preserve different local optima, i.e., fitting the multi-modal property in nature (thereby satisfying \textbf{Goal 1}).
    \item The global optimum of the original $f_t(\vect{x})$ is still Pareto-optimal in the transformed space. In particular, if adaptation plan $\vect{x_1}$ has a better target performance objective than $\vect{x_2}$ 
(i.e., $f_t(\vect{x_1}) < f_t(\vect{x_2})$), 
then whatever their auxiliary objective values are,
$\vect{x_2}$ will not be better than $\vect{x_1}$ on both $g_1$ and $g_2$;
in the best case for $\vect{x_2}$, 
they are nondominated to each other (hence suitable for \textbf{Goal 2}). Note that naively optimize the raw $f_t(\vect{x})$ and $f_a(\vect{x})$ (instead of $g_1(\vect{x})$ and $g_2(\vect{x})$) is harmful to \textbf{Goal 2}, as the search would be forced to optimize $f_a(\vect{x})$, which is of no interest to us.
\end{enumerate}

In this regard, MMO is a perfect fit for our problem of lifelong dynamic optimization for SAS planning. We refer interested readers to the work of Chen and Li~\cite{ChenMMO21} for more detailed examples and proofs of the above. 

However, we cannot directly apply MMO to our problem due primarily to two reasons:

\begin{enumerate}
    \item The $f_a(\vect{x})$ in MMO was deigned to be an additional performance objective. This is, however, not ideal for SAS planning, as it may introduce extra measurement overhead that can be harmful for runtime adaptation.
    \item More importantly, since $f_t(\vect{x})$ and $f_a(\vect{x})$ often come with rather different scales for SASs~\cite{Chen2018FEMOSAA,DBLP:journals/ase/GerasimouCT18}, normalization is required. Chen and Li~\cite{ChenMMO21} has shown that under such a case, the $w$ can become a highly-sensitive parameter to tune before using MMO in the search. This is clearly difficult for SASs where pre-tuning is infeasible.
\end{enumerate}

Thus in this work, we refine MMO by replacing the $f_a(\vect{x})$.

\begin{algorithm}[t!]
	\DontPrintSemicolon
	\footnotesize
	
	\caption{Multi-objectivization Planner for SASs}
	\label{alg:mmo}
	\KwIn{Adaptation plan space $\mathcal{V}$; the Cyber-Twin of managed system $\mathcal{F}$; adaptation interval $k$}

	\kwDeclare{$s_{best}$ the best adaptation plan on $f_t(\vect{x})$}
	Randomly initialize a population of $n$ adaptation plan $\mathcal{P}$\\

	\textsc{measure($\mathcal{P},\mathcal{F}$)}\\
    $t = t + n$\\
	\While{The search space has not been covered in the current environment and the system still runs}
	{  
		
		$\mathcal{P'}=\emptyset$\\
		
		\While{$\mathcal{P'}<n$}
		{ 
	
			$\{s_x,s_y\}\leftarrow$\textsc{mating($\mathcal{P}$)}
			
			$\{o_x,o_y\}\leftarrow$\textsc{doCrossoverAndMutation($\mathcal{V}, s_x,s_y$)}\\
		\tcc{in the same environment, only newly explored plans need to be measured; re-emerged ones can reuse the previously measured objectives.}
		   \If{$o_x$ is new in current environment} {
		   	\textsc{measure($o_x,\mathcal{F}$)}\\
		   	$t = t+1$\\
		   }
		   \If{$o_y$ is new in current environment} {
		   	\textsc{measure($o_y,\mathcal{F}$)}\\
		   	 $t = t+1$\\
		   }

			$\mathcal{P'}\leftarrow\mathcal{P'}\bigcup\{o_x,o_y\}$\\
		}
	 
		$\mathcal{U'}\leftarrow$$\mathcal{P}\bigcup\mathcal{P'}$\\
	
	    \For{$\forall s' \in \mathcal{U'}$} {
	        $\mathcal{N}\leftarrow$\textsc{getCloestNeighboringPlans($s'$,$\mathcal{U'}$)}\\
	        
	        $a\leftarrow$\textsc{getMostDissimilarPlanOnTarget($s'$,$\mathcal{N}$)}\\
	        
	        $f_a(\vect{x})$ of $s'\leftarrow f_t(\vect{x})$ of $a$\\
	    }

		{\textsc{computeMMOModel($\mathcal{U'}$)}}\\
	
		$\mathcal{U}\leftarrow$\textsc{nondominatedSorting($\mathcal{U'}$)}\\
		$\mathcal{P}\leftarrow$top $n$ adaptation plans from $\mathcal{U}$\\
	    $s'_{best}\leftarrow$\textsc{bestPlan($\mathcal{P}$)}\\
	    \If{$s'_{best}$ is better than $s_{best}$} {
	    $s_{best}\leftarrow s'_{best}$\\

	    }
	    
	    \tcc{adapt the managed system with a better plan upon every $k$ new measurements.}
	    \If{$t \geq k$ and $s_{best}$ has not been sent} {
	    $t = 0$\\
	    \textsc{sendForAdaptation($s_{best}$)}
	    }
	}
	
	
\end{algorithm}

\subsection{Refined MMO based Planning}

Refining MMO by finding an appropriate replacement of the auxiliary objective $f_a(\vect{x})$ is non-trivial, as we need to ensure that (1) $f_a(\vect{x})$ is an easy-to-measure objective; and (2) it should be within the same scale as that of $f_t(\vect{x})$ (hence the $w$ can be removed by setting $w=1$) while (3) similarly performing adaptation plans on $f_t(\vect{x})$ can have very different values on $f_a(\vect{x})$, which is an important requirement for multi-objectivization with MMO as shown by Chen and Li~\cite{ChenMMO21}.

Since MMO is often used with a population-based multi-objective search algorithm (we use NSGA-II in this work), to replace the $f_a(\vect{x})$ in MMO, we focus on using the neighboring adaptation plans in the current population (and generated offsprings) during the search. Algorithm~\ref{alg:mmo} shows the pseudo-code for the refined MMO using NSGA-II as the underlying algorithm for SAS planning. As can be seen, to use MMO, the key is that selecting which adaptation plan to preserve will be done in the transformed space ($g_1(\vect{x})$ and $g_2(\vect{x})$) instead of the original space (line 22). Since it is impractical to influence the managed system during planning, the measurements of unique adaptation plans are conducted on a Cyber-Twin of the system to be managed (line 2, 11 and 14).

Specifically, at each iteration/generation, we compute the $f_a(\vect{x_s})$ for an adaptation plan $\vect{x_s}$ as the follows:

\begin{enumerate}
    \item Find the adaptation plan(s) $\mathcal{N} = \{\vect{x_1},\vect{x_2},...,\vect{x_k}\}$ with the closest distance (in terms of the variable space $\vect{x}$) to $\vect{x_s}$ in the current population/offsprings. We use Euclidean distance as it offers good discriminative power (line 19).
    \item Identify the $\vect{x_a}$ from $\mathcal{N}$ such that its $f_t(\vect{x_a})$ has the largest difference to $f_t(\vect{x_s})$, and then set $f_a(\vect{x_s})=f_t(\vect{x_a})$. In this way, we reduce the chance of having $f_t(\vect{x_s})=f_a(\vect{x_s})$, which essentially invalidates multi-objectivization (line 20-21).
\end{enumerate}

Clearly, the new auxiliary objective $f_a(\vect{x})$ is easy-to-measure and it will have the same scale as $f_t(\vect{x})$. While the neighboring plans for SASs may behave similarly on the performance, Nair \textit{et al.}~\cite{nair2018finding} have shown that for highly-configurable systems, those neighboring plans can also lead to radically different results. The is because, for example, when two plans only differ on whether to turn on the cache option, the performance results can differ dramatically even though these plans exhibit very similar representation in the search algorithm. Similar observations have also been registered by the others~\cite{DBLP:conf/mascots/JamshidiC16}. As a result, our design also ensures that adaptation plans can have similarly-performing $f_t(\vect{x})$ but very different $f_a(\vect{x})$.

Since \approach~is a lifelong dynamic optimization framework, the multi-objectivization planner terminates when all the search space in an environment has been explored, or otherwise, it runs throughout the lifetime of the SASs. An adaptation to the managed system is triggered every $k$ new measurements and when a better plan has been found (line 29-31). Here, $k$ serves as a parameter that determines the extent of convergence and tolerated overhead for the planning in \approachfs.

\subsection{Architecture}                                       

\begin{figure}[t!]
\centering
\includegraphics[width=0.8\columnwidth]{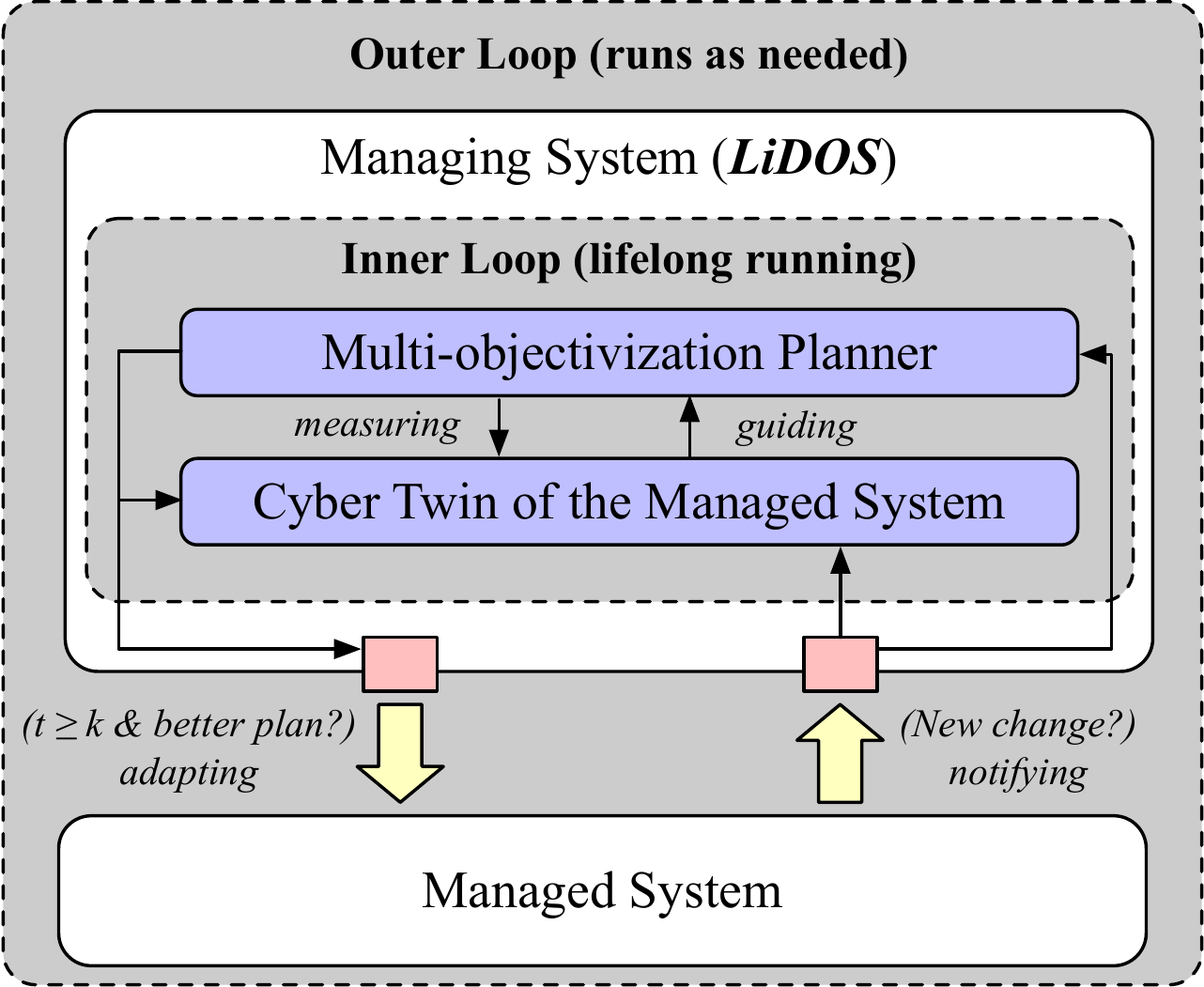}
\caption{\approach~architecture with a hierarchical feedback loop.}
\label{fig:arch}
\end{figure}

Figure~\ref{fig:arch} shows the architecture of \approachd, which uses a hierarchical feedback loop. As can be seen, we adopt the external management style for SAS~\cite{DBLP:journals/taas/SalehieT09} where the managing system and the managed system are separated. 

The multi-objectivization planner, which is the core of \approachd, runs in the inner loop. To enable lifelong optimization, the planning process runs as soon as the managed system is deployed and it will not stop unless the search space in an environment has been covered or the managed system terminates. To achieve dynamic optimization, the planner searches for a better adaptation plan by measuring a Cyber-Twin of the managed system, hence providing accurate measurements of the adaptation plans without influencing the system at production. Such a Cyber-Twin serves as a replica of the managed system (may be deployed on a different machine) and runs under exactly the same environment as what has been experienced by the managed system. In most cases, the Cyber-Twin can simply be another running instance of the managed system.

In contrast, the outer loop follows the normal adaptation process where it notifies the Cyber-Twin in the inner loop whenever there is an environmental change, thereby the Cyber-Twin can be configured under the same environment (e.g., by feeding the same workload to it). The outer loop will also set $t=0$ in the planner and trigger re-measurement of the plans in the current population. In this way, the planner would be aware of the change via measuring the Cyber-Twin. Adaptation to the managed system will occur whenever there are $k$ new measurements and the planner in the inner loop finds a better adaptation plan.

Note that the architecture of \approach~can be easily adopted to other common patterns for SASs, e.g., the MAPE-K loop.

\section{Experiment Setup}
\label{sec:study}

In this work, we seek to understand the following research questions (RQs):

\begin{itemize}
    \item \textbf{RQ1:} Is being ``dynamic'' more beneficial than being ``stationary'' for SASs?
    \item \textbf{RQ2:} How effective is the adaptation achieved by \approach?
    \item \textbf{RQ3:} How efficient for \approach~to produce a promising adaptation plan?
\end{itemize}

We ask \textbf{RQ1} to verify whether considering lifelong dynamic optimization for SASs is more necessary than its stationary counterpart where an optimization run is simply restarted in the planning. We study \textbf{RQ2} to confirm the effectiveness of \approach~against state-of-the-art planners. Yet, it would still be meaningless if \approach~requires a large amount of planning overhead to become effective. Therefore, \textbf{RQ3} seeks to understand how much speedup that \approach~can achieve against the state-of-the-arts.

\subsection{Subject SASs}

We study three real-world SASs that are widely used in prior work of configurable and self-adaptive systems:

\begin{itemize}
    \item \textsc{Apache Storm:} a distributed stream processing computation framework that executes a series of job batches. Adaptation means altering runtime options such as \texttt{message\_size}, \texttt{num\_counters}, and \texttt{num\_splitters}. The changing environments are related to the workload and type of a given batch of jobs. 
    \item \textsc{Keras/DNN:} the deep neural network (DNN) from the well-known \textsc{Keras} framework for classification tasks as part of a learning system. When deployed and trained, the learning system then classifies newly given samples. Adaptation refers to adjust options includes \texttt{batch\_size}, \texttt{num\_filters}, and \texttt{log\_decay}, for (re-)training the network in the system. The environment can change depending on the given dataset for training.
    \item \textsc{x264:} A system for encoding videos. Adaptation means tuning runtime options includes \texttt{no\_mbtree}, \texttt{bframes}, and \texttt{ip\_ratio}. Environment is determined by the sizes and frames of different videos.
\end{itemize}

More detailed information about the SASs has been shown in Table~\ref{tb:sys}, where we consider two environments that can change at runtime for each SAS and use the same setting as previous work. Note that since the measurement of an adaptation plan can be expensive (even with a Cyber-Twin)~\cite{ChenMMO21}, it is important to find an effective plan using as few measurements as possible. To expedite the experiment, we use the readily available dataset of those systems~\cite{nair2018finding,DBLP:conf/mascots/JamshidiC16,DBLP:conf/sigsoft/JamshidiVKS18,DBLP:journals/corr/abs-2106-02716} from existing work to emulate the Cyber-Twin of the managed system.

\begin{table}[t!]
\caption{Real-world self-adaptive systems studied.}
\label{tb:sys}
\setlength{\tabcolsep}{0.9mm}
\centering
\footnotesize
\begin{tabular}{lllccl}\toprule

\textbf{SAS}&\textbf{$\lvert \mathbfcal{P} \rvert$}&\textbf{Environment}&\textbf{$\lvert \mathbfcal{O} \rvert$}&\textbf{$\lvert \mathbfcal{S} \rvert$}&\textbf{Used By}\\

\midrule

\multirow{2}{*}{\textsc{Storm}}&\multirow{2}{*}{Throughput}&\textsc{WordCount}&\multirow{2}{*}{12}&\multirow{2}{*}{1,914}&\multirow{2}{*}{\cite{nair2018finding,DBLP:conf/mascots/JamshidiC16,DBLP:journals/corr/abs-2106-02716}}\\

&&\textsc{RollingCount}&&&\\ \hline

\multirow{2}{*}{\textsc{Keras/DNN}}&\multirow{2}{*}{AUC}&\textsc{ShapesAll}&\multirow{2}{*}{12}&\multirow{2}{*}{16,384}&\multirow{2}{*}{\cite{DBLP:conf/mascots/MendesCRG20,DBLP:conf/sigsoft/JamshidiVKS18}}\\

&&\textsc{Adiac}&&&\\ \hline

\multirow{2}{*}{\textsc{x264}}&\multirow{2}{*}{Latency}&\textsc{128/44}&\multirow{2}{*}{17}&\multirow{2}{*}{53,662}&\multirow{2}{*}{\cite{nair2018finding,DBLP:conf/icse/SiegmundKKABRS12,DBLP:journals/corr/abs-2106-02716}}\\

&&\textsc{8/2}&&&\\


\bottomrule
\end{tabular}
\centering
 \begin{tablenotes}
    \footnotesize
    \item $\lvert \mathbfcal{P} \rvert$, $\lvert \mathbfcal{O} \rvert$ and $\lvert \mathbfcal{S} \rvert$ denote performance objective, $\#$ adaptation options and search space, respectively. 
    \end{tablenotes}
\end{table}

\subsection{Compared SAS Planners}
\label{sec:compared}

To answer our RQs, we compare \approach~with the following variants and state-of-the-art planners:

\begin{itemize}
    \item \approachs~: This is a stationary variant of the \approachd, such that a simple restart is triggered whenever a change of environment is detected. All other parts are identical.
    \item \textbf{Pseudo-dynamic planner} (Ramirez \textit{et al.}~\cite{DBLP:conf/icac/RamirezKCM09} \& Kinneer \textit{et al.}~\cite{DBLP:journals/taas/KinneerGG21}): The planner uses a vanilla search algorithm that runs continuously over environmental changes, i.e., single-objective genetic algorithm (SOGA)~\cite{whitley1994genetic}, without tinkering how dynamics are handled during the search. 
    \item \textbf{Stationary planner} (Chen \textit{et al.}~\cite{Chen2018FEMOSAA}): The planner where a new search run is triggered (with random initial adaptation plans) when an environment change is detected. This is similar to \approachs~but without the multi-objectivization/NSGA-II and SOGA is used instead.
\end{itemize}

All planners are implemented in Java with search algorithms from jMetal~\cite{DBLP:journals/aes/DurilloN11}. Experiments are conducted on a machine with quad-core CPU at 2.8GHz and 16GB RAM.

\subsection{Parameter Settings}

For the key parameters of the stochastic search algorithm in all the planners, we apply the binary tournament for mating selection, together with the boundary mutation and uniformed crossover, as used in prior work for SASs~\cite{Chen2018FEMOSAA,DBLP:conf/sigsoft/ShahbazianKBM20,DBLP:journals/infsof/ChenLY19}. The mutation and crossover rates are set to 0.1 and 0.9, respectively, with a population size of 20, which is widely used~\cite{Chen2018FEMOSAA,ChenMMO21}. 


To provide fair statistics for \textbf{RQ1} and \textbf{RQ2} (especially for those stationary counterparts), we set the adaptation interval $k$ as 150 measurements (of the Cyber-Twin), as it is the smallest number that there is no change on the best adaptation plan in the last 5 iterations (for all planners), implying a reasonable convergence. Clearer trajectories of the performance change over time will be demonstrated for \textbf{RQ3}.

\subsection{Statistical Validation}

To verify statistical significance, all experiments in this work are repeated 50 runs.

\subsubsection{Pair-wise Comparisons}

We use the following methods for comparing the target performance objective of two planners: 


\begin{itemize}

    \item \textbf{Non-parametric test:} To verify statistical significance, we leverage the Wilcoxon rank-sum test~\cite{Wilcoxon1945IndividualCB} --- a widely used non-parametric test for SBSE and has been recommended in software engineering research for its strong statistical power on pairwise comparisons~\cite{ArcuriB11}. The standard $a=0.05$ is set as the significance level over 50 runs. If the $p<0.05$, we say the magnitude of differences in the comparisons are significant. 
    
    
    \item \textbf{Effect size:} To ensure the resulted differences are not generated from a trivial effect, we use $\hat{A}_{12}$~\cite{Vargha2000ACA} to verify the effect size over 50 runs. According to Vargha and Delaney~\cite{Vargha2000ACA}, when comparing \approach~and its counterpart in this work, $\hat{A}_{12}>0.5$ denotes that the \approach~is better for more than 50\% of the times. In particular, $0.56\leq \hat{A}_{12}<0.64$ indicates a small effect size while $0.64 \leq \hat{A}_{12} < 0.71$ and $\hat{A}_{12} \geq 0.71$ mean a medium and a large effect size, respectively.

\end{itemize}

As such, we say a comparison is statistically significant only if it has $\hat{A}_{12} \geq 0.56$ (or $\hat{A}_{12} \leq 0.44$) and $p < 0.05$.

\subsubsection{Three or More Comparisons}

In case more than two planners need to be compared, we apply Scott-Knott test~\cite{DBLP:journals/tse/MittasA13} on all comparisons on the target performance objective over 50 runs, as recommended by Mittas and Angelis~\cite{DBLP:journals/tse/MittasA13}. In a nutshell, Scott-Knott sorts the list of treatments (the planners) by their median values of the target performance objective. Next, it splits the list into two sub-lists with the largest expected difference~\cite{xia2018hyperparameter}. For example, suppose that we compare $A$, $B$, and $C$, a possible split could be: $\{A, B\}$, $\{C\}$, with the rank of 1 and 2, respectively. This means that, in the statistical sense, $A$ and $B$ perform similarly, but they are significantly better than $C$. Formally, Scott-Knott test aims to find the best split by maximizing the difference $\Delta$ in the expected mean before and after each split:
\begin{equation}
    \Delta = \frac{|l_1|}{|l|}(\overline{l_1} - \overline{l})^2 + \frac{|l_2|}{|l|}(\overline{l_2} - \overline{l})^2
\end{equation}
whereby $|l_1|$ and $|l_2|$ are the sizes of two sub-lists ($l_1$ and $l_2$) from list $l$ with a size $|l|$. $\overline{l_1}$, $\overline{l_2}$, and $\overline{l}$ denote their mean values of the target performance objective.

During the splitting, we apply a statistical hypothesis test $H$ to check if $l_1$ and $l_2$ are significantly different. This is done by using bootstrapping and $\hat{A}_{12}$~\cite{Vargha2000ACA}. If that is the case, Scott-Knott recurses on the splits. In other words, we divide the planners into different sub-lists if both bootstrap sampling and effect size test suggest that a split is statistically significant (with a confidence level of 99\%) and with a good effect $\hat{A}_{12} \geq 0.6$. The sub-lists are then ranked based on their mean values of the target performance objective.
\section{Results}
\label{sec:results}

\subsection{RQ1: Benefit of Being ``Dynamic''}

\subsubsection{Method}

To answer \textbf{RQ1}, we compare \approach~and \approachs~over changing environments under reasonable convergence of the planning. For each SAS, we run the system transits from one environment to another and set $k=150$. That is, for all planners, the planning is allowed to execute for 150 measurements and trigger an (better) adaptation\footnote{The task under an environment will not be processed until the first adaptation has been taken.}. After the task in the current environment has been completed, the environment would be changed to another\footnote{While the planner runs continuously, we observed no further new measurements when the environmental change took place for all cases.} and the planner runs for a further 150 measurements before triggering the next adaptation. We report the target performance objective achieved by the last measurement. Wilcoxon rank-sum test and $\hat{A}_{12}$ are used for pair-wise comparisons over 50 runs. 

\subsubsection{Result}

As we can see from Table~\ref{tb:rq1}, \approach~wins all the 6 cases from which 4 of them have $p<0.05$ and at least medium effect size. The improvement in terms of median and IQR of the target performance objective has also been considerably high. Remarkably, it achieves almost $10\times$ median improvement on the case of \textsc{x264} when the environment changes from the video with \textsc{128/44} to one with \textsc{8/2}. 

Since \approach~and \approachs~share the same optmization model and search algorithm, the above are clear evidence on the benefits of being dynamic when planning for SASs, as the stationary counterpart, which restarts the search upon environmental change, will waste the accumulated information that can be useful after the change. Therefore, we say:

\begin{quotebox}
   \noindent
   \textit{\textbf{RQ1:} Explicitly considering ``dynamic'' in the planning of SASs is generally more beneficial than its stationary counterpart with up to $10\times$ improvement.}
\end{quotebox}


\begin{table}[t!]
\caption{Comparing \approach~and the variant of \approach~with a stationary setting (denotes as \approachs~) over 50 runs. \quartexp{0}{20}{10}{20} visualizes the 25th, 50th, and 75th percentile of the results (normalized in the scale of all values found). $\hat{A}_{12}>0.5$ means \approach~wins (highlighted in \setlength{\fboxsep}{1.5pt}\colorbox{steel!30}{blue}) while $\hat{A}_{12}<0.5$ denotes that it loses. Statistically significant comparisons are shown in bold. $\uparrow$ and $\downarrow$ denote maximal and minimal value of the performance objective is preferred, respectively; $\rightarrow$ indicates the direction of environment change at runtime.}
    \label{tb:rq1}
    \footnotesize
  \begin{center}
    \begin{adjustbox}{max width = 1\columnwidth}

    \begin{tabular}{lccllc} \toprule
    \textbf{Method}&\textbf{Median}&\textbf{IQR}&\textbf{$\hat{A}_{12}$ ($p$ value)}&\\ \midrule
    \approachs~&4803&206&\cellcolor{steel!30}&\quart{49.64292997376618}{50.35707002623382}{71.74046107182978}{100}&\multirow{2}{*}{$\uparrow$}  \\
    \approach~&4883&122&\cellcolor{steel!30}\multirow{-2}{*}{0.56 ($p=0.347$)}&\quart{61.37522957651485}{29.78983862086094}{91.1650681973758}{100}&\\ 
    
    &&&&&\\ [-0.2cm]
    
    \rowcolor{black!20} \multicolumn{6}{c}{Throughput (msgs/min) for \textsc{Storm}, \textsc{WordCount} $\rightarrow$ \textsc{RollingCount}} \\ \midrule
    
     \textbf{Method}&\textbf{Median}&\textbf{IQR}&\textbf{$\hat{A}_{12}$ ($p$ value)}&\\ \midrule
    
     \approachs~&10119&56&\cellcolor{steel!30}&\quart{33.63506369261786}{32.06288640544578}{53.64459280208074}{100}&\multirow{2}{*}{$\uparrow$}   \\
    \approach~&10200&60&\cellcolor{steel!30}\multirow{-2}{*}{\textbf{0.75 ($p<0.001$)}}&\quart{65.69795009806364}{34.30204990193636}{100.0}{100}&  \\ 
    
    &&&&&\\ [-0.2cm]
    
    \rowcolor{black!20} \multicolumn{6}{c}{Throughput (msgs/min) for \textsc{Storm}, \textsc{RollingCount} $\rightarrow$ \textsc{WordCount}} \\ \midrule
    
     \textbf{Method}&\textbf{Median}&\textbf{IQR}&\textbf{$\hat{A}_{12}$ ($p$ value)}&\\ \midrule
    
      \approachs~&0.064&0.105&\cellcolor{steel!30}&\quart{3.4188034045218862}{35.04273497989625}{7.692307743721216}{100}&\multirow{2}{*}{$\uparrow$}   \\
    \approach~&0.274&0.225&\cellcolor{steel!30}\multirow{-2}{*}{\textbf{0.70 ($p=0.038$)}}&\quart{5.128205273876829}{75.21367523366938}{77.77777770351378}{100}&  \\ 
    
       &&&&&\\ [-0.2cm]
    
    \rowcolor{black!20} \multicolumn{6}{c}{AUC for \textsc{Keras/DNN}, \textsc{ShapesAll} $\rightarrow$ \textsc{Adaic}} \\ \midrule
    
     \textbf{Method}&\textbf{Median}&\textbf{IQR}&\textbf{$\hat{A}_{12}$ ($p$ value)}&\\ \midrule
    
      \approachs~&0.178&0.278&\cellcolor{steel!30}&\quart{3.571428640306118}{59.64285702882653}{33.928571332908156}{100}&\multirow{2}{*}{$\uparrow$}   \\
    \approach~&0.292&0.292&\cellcolor{steel!30}\multirow{-2}{*}{0.58 ($p=0.167$)}&\quart{2.857142783673472}{62.500000026785706}{58.21428574413265}{100}&  \\ 
    
       &&&&&\\ [-0.2cm]
    
    \rowcolor{black!20} \multicolumn{6}{c}{AUC for \textsc{Keras/DNN}, \textsc{Adaic} $\rightarrow$ \textsc{ShapesAll}} \\ \midrule
    
     \textbf{Method}&\textbf{Median}&\textbf{IQR}&\textbf{$\hat{A}_{12}$ ($p$ value)}&\\ \midrule
    
      \approachs~&31.147&32.210&\cellcolor{steel!30}&\quart{1.6071034326998468}{81.73819922195182}{70.19258341940248}{100}&\multirow{2}{*}{$\downarrow$}   \\
    \approach~&3.877&0.703&\cellcolor{steel!30}\multirow{-2}{*}{\textbf{0.80 ($p<0.001$)}}&\quart{0.4651540489718638}{1.7847400035019791}{0.9896894658975841}{100}&  \\ 
    
       &&&&&\\ [-0.2cm]
    
    \rowcolor{black!20} \multicolumn{6}{c}{Latency (s) for \textsc{x264}, \textsc{128/44} $\rightarrow$ \textsc{8/2}} \\ \midrule
    
     \textbf{Method}&\textbf{Median}&\textbf{IQR}&\textbf{$\hat{A}_{12}$ ($p$ value)}&\\ \midrule
    
      \approachs~&113.350&13.610&\cellcolor{steel!30}&\quart{21.997550336291578}{29.2455465543546}{33.837591593785596}{100}&\multirow{2}{*}{$\downarrow$}   \\
    \approach~&106.510&9.050&\cellcolor{steel!30}\multirow{-2}{*}{\textbf{0.70 ($p<0.001$)}}&\quart{11.683176827040876}{19.446891720566427}{19.139609343103363}{100}&  \\ 
    
       &&&&&\\ [-0.2cm]
    
    \rowcolor{black!20} \multicolumn{6}{c}{Latency (s) for \textsc{x264}, \textsc{8/2} $\rightarrow$ \textsc{128/44}} \\ \midrule
    
    \end{tabular}
  \end{adjustbox}
   \end{center}
\end{table}

\subsection{RQ2: Effectiveness of Adaptation}

\subsubsection{Method}

For \textbf{RQ2}, we compare \approach~with the state-of-the-art pseudo-dynamic planner (Ramirez \textit{et al.}~\cite{DBLP:conf/icac/RamirezKCM09} \& Kinneer \textit{et al.}~\cite{DBLP:journals/taas/KinneerGG21}) and the stationary planner (Chen \textit{et al.}~\cite{Chen2018FEMOSAA}) discussed in Section~\ref{sec:compared}. We use the same setting for environmental changes as that for \textbf{RQ1} and report the results by the last measurement. Since there are more than two planners, we use Scott-Knott test to rank the results over 50 runs.


\subsubsection{Result}

The results are reported in Table~\ref{tb:rq2}. Clearly, \approach~has been constantly ranked as the first with considerable improvement on the median performance over the others. Overall, the pseudo-dynamic planner by Ramirez \textit{et al.}~\cite{DBLP:conf/icac/RamirezKCM09} \& Kinneer \textit{et al.}~\cite{DBLP:journals/taas/KinneerGG21} tends to be better than the stationary one used by Chen \textit{et al.}~\cite{Chen2018FEMOSAA}, but the improvement is not significant. This is not surprising, as the SOGA may be trapped at some undesired local optima, which tends to be not useful but misleading for the search after the environmental change.

\begin{figure}[t!]
\centering
\begin{subfigure}[h]{\columnwidth}
\includegraphics[width=\columnwidth]{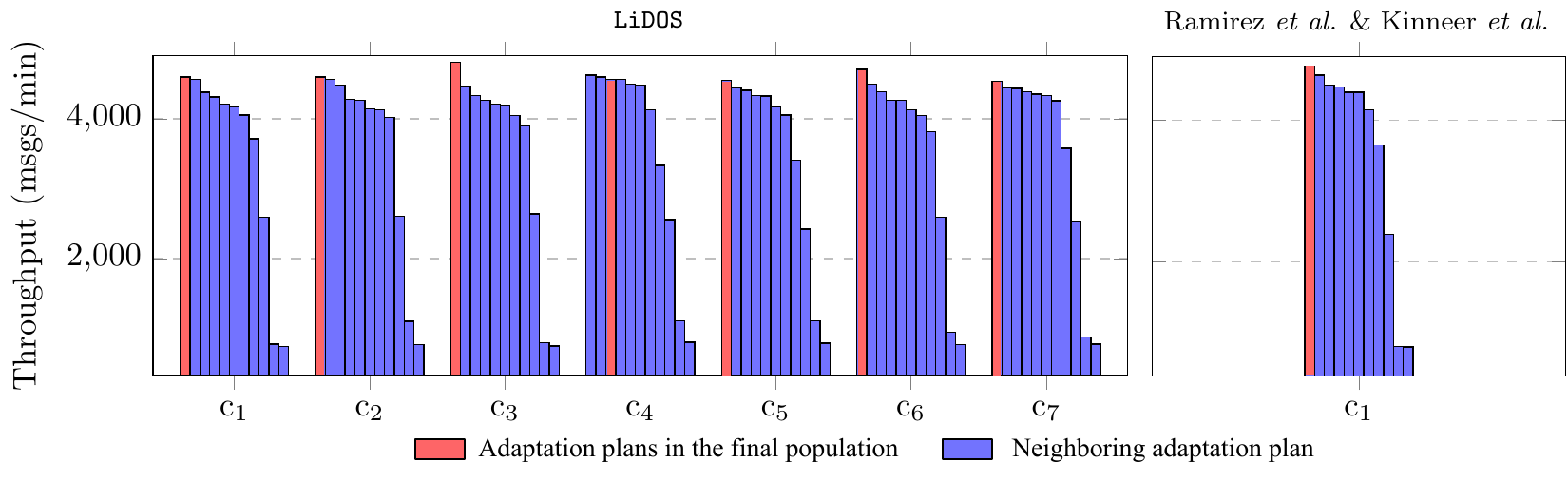}
\subcaption{\footnotesize Descendingly sorted performance of the preserved adaptation plans and their neighbouring plans before the change (\textsc{RollingCount}).}
\end{subfigure}

\begin{subfigure}[h]{\columnwidth}
\includegraphics[width=\columnwidth]{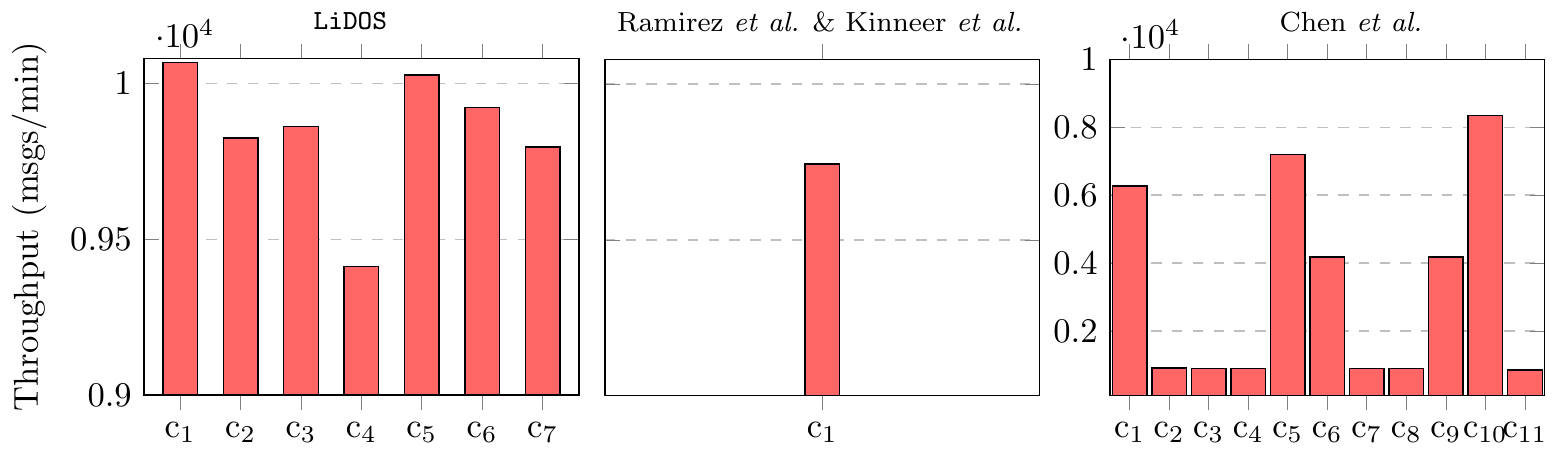}
\subcaption{\footnotesize Performance of the same preserved adaptation plans after the change (\textsc{WordCount}).}
\end{subfigure}

\caption{A run on the performance of preserved adaptation plans before and after the runtime environment change for \textsc{Storm}. (a) illustrates the performance of all plans found before the change and whether they are local optima with respect to their neighboring configurations; the x-axis indicates the unique plans. (b) shows the performance of the same unique plans after the change. The rightmost figure in (b) shows the result of the stationary planner by Chen \textit{et al.}~\cite{Chen2018FEMOSAA}, which essentially are some randomly generated adaptation plans.}
\label{fig:details}
\end{figure}

To confirm the effectiveness of preserving different local optima for dynamic optimization in SAS planning, Figure~\ref{fig:details} plots the performance of the adaptation plans found before and after the environmental change for one run of \textsc{Storm}. From Figure~\ref{fig:details}a, we see that the plans preserved by \approach~right before the change occur contain a wide range of local optima (In fact, $c_3$ is the global optimum) --- 6 out of 7 are the best among all their neighboring plans. In contrast, the pseudo-dynamic planner by Ramirez \textit{et al.}~\cite{DBLP:conf/icac/RamirezKCM09} \& Kinneer \textit{et al.}~\cite{DBLP:journals/taas/KinneerGG21} is more tempting to converge to one adaptation plan, which can be a local optimum. When re-measuring the same adaptation plans after the change (Figure~\ref{fig:details}b), we observe that:

\begin{table}[t!]
\caption{Comparing \approach~and the state-of-the-art pseudo-dynamic and stationary planners for SASs over 50 runs. In each case, the planners are sorted by Scott-Knott rank, median, and then IQR. The same formats as Table~\ref{tb:rq1} applied.}
    \label{tb:rq2}
    \footnotesize
    \setlength{\tabcolsep}{1mm}
  \begin{center}
    \begin{adjustbox}{max width = 1\columnwidth}

    \begin{tabular}{lccllc} \toprule
    \textbf{Method}&\textbf{Rank}&\textbf{Median}&\textbf{IQR}&\\ \midrule
     \approach~&1&4883&122&\quart{56.01765803344089}{33.921932868102026}{89.93959090154291}{100}&\multirow{3}{*}{$\uparrow$}\\ 
    
    Ramirez \textit{et al.}~\cite{DBLP:conf/icac/RamirezKCM09} \& Kinneer \textit{et al.}~\cite{DBLP:journals/taas/KinneerGG21}&2&4761&0&\quart{56.01765803344089}{0.0}{56.01765803344089}{100}&\\ 
    
    Chen \textit{et al.}~\cite{Chen2018FEMOSAA}&2&4761&244&\quart{21.979553789981686}{67.96003711156122}{56.01765803344089}{100}&\\
    
     &&&&&\\ [-0.2cm]
    
    \rowcolor{black!20} \multicolumn{6}{c}{Throughput (msgs/min) for \textsc{Storm}, \textsc{WordCount} $\rightarrow$ \textsc{RollingCount}} \\ \midrule
    
      \textbf{Method}&\textbf{Rank}&\textbf{Median}&\textbf{IQR}&\\ \midrule
    
        \approach~&1&10200&60&\quart{65.20057864477809}{34.79942135522192}{100.0}{100}&\multirow{3}{*}{$\uparrow$}\\ 
    
    Ramirez \textit{et al.}~\cite{DBLP:conf/icac/RamirezKCM09} \& Kinneer \textit{et al.}~\cite{DBLP:journals/taas/KinneerGG21}&2&10140&21&\quart{52.97245057409188}{12.228128070686212}{65.20057864477809}{100}&\\ 
    
    Chen \textit{et al.}~\cite{Chen2018FEMOSAA}&2&10119&116&\quart{32.67278811687909}{67.32721188312091}{52.97245057409188}{100}&\\
    
      &&&&&\\ [-0.2cm]
    
    \rowcolor{black!20} \multicolumn{6}{c}{Throughput (msgs/min) for \textsc{Storm}, \textsc{RollingCount} $\rightarrow$ \textsc{WordCount}} \\ \midrule
    
      \textbf{Method}&\textbf{Rank}&\textbf{Median}&\textbf{IQR}&\\ \midrule
    
        \approach~&1&0.274&0.225&\quart{5.128205273876829}{75.21367523366938}{77.77777770351378}{100}&\multirow{3}{*}{$\uparrow$}\\ 
    
    Ramirez \textit{et al.}~\cite{DBLP:conf/icac/RamirezKCM09} \& Kinneer \textit{et al.}~\cite{DBLP:journals/taas/KinneerGG21}&1&0.174&0.233&\quart{0.0}{77.77777770351378}{44.44444459297245}{100}&\\ 
    
    Chen \textit{et al.}~\cite{Chen2018FEMOSAA}&2&0.059&0.113&\quart{0.854700934677477}{37.60683744974066}{5.982905874366262}{100}&\\
    
      &&&&&\\ [-0.2cm]
    
    \rowcolor{black!20} \multicolumn{6}{c}{AUC for \textsc{Keras/DNN}, \textsc{ShapesAll} $\rightarrow$ \textsc{Adaic}} \\ \midrule
    
      \textbf{Method}&\textbf{Rank}&\textbf{Median}&\textbf{IQR}&\\ \midrule
    
         \approach~&1&0.292&0.292&\quart{2.1582733781895347}{62.949640269137205}{57.91366912530407}{100}&\multirow{3}{*}{$\uparrow$}\\

    Chen \textit{et al.}~\cite{Chen2018FEMOSAA}&2&0.163&0.258&\quart{2.1582733781895347}{55.755395747114534}{30.215827294653486}{100}&\\
    
     Ramirez \textit{et al.}~\cite{DBLP:conf/icac/RamirezKCM09} \& Kinneer \textit{et al.}~\cite{DBLP:journals/taas/KinneerGG21}&2&0.112&0.245&\quart{2.1582733781895347}{52.8776977656436}{19.064748317892445}{100}&\\

     &&&&&\\ [-0.2cm]
    
    \rowcolor{black!20} \multicolumn{6}{c}{AUC for \textsc{Keras/DNN}, \textsc{Adaic} $\rightarrow$ \textsc{ShapesAll}} \\ \midrule
    
      \textbf{Method}&\textbf{Rank}&\textbf{Median}&\textbf{IQR}&\\ \midrule
    
        \approach~&1&3.877&0.703&\quart{0.43017768004449525}{1.6505398929366852}{0.9152716596691403}{100}&\multirow{3}{*}{$\downarrow$}\\ 
    
    Ramirez \textit{et al.}~\cite{DBLP:conf/icac/RamirezKCM09} \& Kinneer \textit{et al.}~\cite{DBLP:journals/taas/KinneerGG21}&1&4.190&26.247&\quart{0.9464847701142676}{61.59778269573318}{1.6505398929366846}{100}&\\ 
    
    Chen \textit{et al.}~\cite{Chen2018FEMOSAA}&2&6.663&32.307&\quart{1.5174734747232472}{75.81899212162317}{2.761304191709514}{100}&\\
    
     &&&&&\\ [-0.2cm]
    
    \rowcolor{black!20} \multicolumn{6}{c}{Latency (s) for \textsc{x264}, \textsc{128/44} $\rightarrow$ \textsc{8/2}} \\ \midrule
    
      \textbf{Method}&\textbf{Rank}&\textbf{Median}&\textbf{IQR}&\\ \midrule
    
      \approach~&1&106.510&9.050&\quart{9.470970439145074}{15.764628007037466}{15.515529465047836}{100}&\multirow{3}{*}{$\downarrow$}\\ 
    
    Ramirez \textit{et al.}~\cite{DBLP:conf/icac/RamirezKCM09} \& Kinneer \textit{et al.}~\cite{DBLP:journals/taas/KinneerGG21}&2&113.310&9.550&\quart{18.772971937220202}{16.635601930078234}{27.36077481840195}{100}&\\ 
    
    Chen \textit{et al.}~\cite{Chen2018FEMOSAA}&3&120.660&13.680&\quart{26.054313933840834}{23.82984653439475}{40.16409148710088}{100}&\\
    
     &&&&&\\ [-0.2cm]
    
    \rowcolor{black!20} \multicolumn{6}{c}{Latency (s) for \textsc{x264}, \textsc{8/2} $\rightarrow$ \textsc{128/44}} \\ \midrule
    
    \end{tabular}
  \end{adjustbox}
   \end{center}
\end{table}

\begin{enumerate}
    \item \approach~immediately obtain the best adaptation plans among those found by the state-of-the-arts, i.e., $c_1$ under \approachd, which is merely one of the local optima that it found before the change in Figure~\ref{fig:details}a. This shows the usefulness of local optima for coping with the dynamic changes that occur in the planning of SASs.
    \item Although the single plan found by Ramirez \textit{et al.}~\cite{DBLP:conf/icac/RamirezKCM09} \& Kinneer \textit{et al.}~\cite{DBLP:journals/taas/KinneerGG21} is promising before (the 2nd best in Figure~\ref{fig:details}a), it can be significantly worsen after the change (only the 7th best in Figure~\ref{fig:details}b). This reveals the importance of exploring and preserving a wide range of local optima.
    \item The random plans in the stationary planner (Chen \textit{et al.}~\cite{Chen2018FEMOSAA}) are of generally worse performance than the other two.
\end{enumerate}

As a result, we conclude that:

\begin{quotebox}
   \noindent
   \textit{\textbf{RQ2:} \approach~is effective as it is generally ranked better (in the statistical sense) than the state-of-the-art with considerable improvement because multiple local optima can be preserved upon environmental change during the lifelong dynamic optimization.}
\end{quotebox}

\begin{figure*}[t!]
\centering
\includegraphics[width=\textwidth]{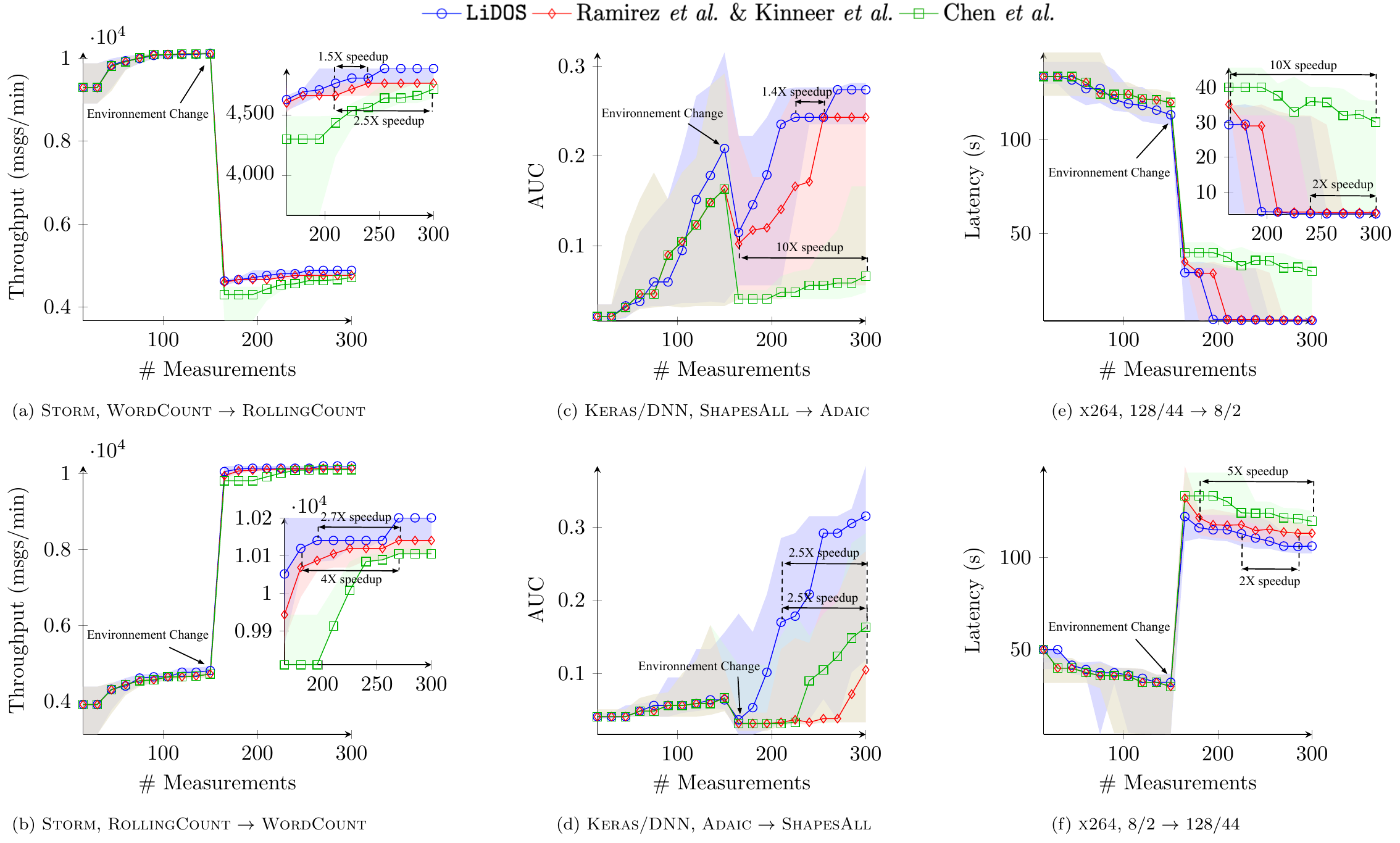}
\caption{Trajectories of adaptation planning with environment change at runtime. The points and highlighted areas denote the median and IQR of the performance objective every 15 measurements over 50 runs, respectively. The small figures are ``zoom-in'' of the differences after the change in greater detail.}
\label{fig:efficiency}
\end{figure*}

\subsection{RQ3: Efficiency of Adaptation}

\subsubsection{Method}

\textbf{RQ1} and \textbf{RQ2} have demonstrated the effectiveness and benefit of \approach~under $k=150$ that leads to a reasonable convergence in the SAS planning. However, the results are less promising if \approach~requires a generally higher planning overhead to achieve such. In \textbf{RQ3}, we investigate the trajectories of the target performance objective when planning across all measurements with the environmental change. We are particularly interested in the efficiency of adaptation after an environmental change: the ``speedup'' of \approach~to reach the best performance as that achieved by the state-of-the-art. To this end, we use the metric of $T_{base} \over T_{LiDOS}$ proposed by Gao \textit{et al.}~\cite{DBLP:conf/icse/GaoZ0LY21}, whereby $T_{base}$ is the smallest number of measurements (after change) for a state-of-the-art planner to obtain its best performance; $T_{LiDOS}$ is the smallest number of measurements (after change) for \approach~to perform the same.

\subsubsection{Result}

Figure~\ref{fig:efficiency} illustrates the results, from which we see that the environmental change indeed causes a severer disruption for the planning. However, thanks to the ability to preserve multiple local optima, \approach~generally starts planning with some better adaptation plans than the state-of-the-art after an environmental change, hence leading to faster and better improvement throughout the planning. It is worth noting that such a benefit can be achieved even if the plans found before the change are not quite promising: while the \approach~generally can better mitigate the issue of being trapped by local optima (thanks to MMO), there are still cases (e.g., Figure~\ref{fig:efficiency}f) where its median is worse than the others. However, even in this case, the multi-objectivization planner in \approach~can still lead to a better trajectory after an environmental change occurs.

When comparing the efficiency, we see that \approach~achieves significant improvement over the others: $1.4\times$ to $5\times$ speedup over the pseudo-dynamic planner by Ramirez \textit{et al.}~\cite{DBLP:conf/icac/RamirezKCM09} \& Kinneer \textit{et al.}~\cite{DBLP:journals/taas/KinneerGG21}; $2.5\times$ to $10\times$ speedup on the stationary planner by Chen \textit{et al.}~\cite{Chen2018FEMOSAA}.

In summary, we state that:

\begin{quotebox}
   \noindent
   \textit{\textbf{RQ3:} \approach~is more efficient than the state-of-the-art planners, achieving up to $5\times$ and $10\times$ speedup over the pseudo-dynamic and stationary one, respectively.}
\end{quotebox}

\section{Threats to Validity}
\label{sec:discussion}

Threats to \textbf{internal validity} can be related to the $k$, which determines the time allowed for planning. We set $k=150$ in this work since we found that it reaches reasonable convergence for the SASs studied, i.e., for all planners, the best adaptation plan found does not change for the last 5 consecutive iterations. Further, in \textbf{RQ3}, we examine the trajectory of planning, which also indicates what would happen if a smaller $k$ is used. The other parameter settings follow what has been pragmatically used from the literature~\cite{Chen2018FEMOSAA,DBLP:conf/sigsoft/ShahbazianKBM20,DBLP:journals/infsof/ChenLY19,ChenMMO21}. However, we acknowledge that examining alternative parameters (and larger $k$) can be an interesting topic and we leave this for future work. To mitigate bias, we repeated 50 runs for each case.

The metrics and evaluation used may possess threats to \textbf{construct validity}. Since we focus on the scenario where there is only a single performance concern, we directly use the measured performance objective in the comparisons (for \textbf{RQ1} and \textbf{RQ2}) and the formula of speedup used by Gao \textit{et al.}~\cite{DBLP:conf/icse/GaoZ0LY21} to evaluate the efficiency of planning (for \textbf{RQ3}). To verify statistical significance and effect size, we use Wilcoxon rank-sum test and $\hat{A}_{12}$ to examine the pairwise comparisons. When comparing more than two planners, we use Scott-Knott test for ranking. Yet, admittedly, measurement noises are possible.

The subject SASs and environments studied may contribute to the threats to \textbf{external validity}. We mitigated this by using 6 systems/environments that are of different domains, scales and performance attributes, as used by prior work~\cite{nair2018finding,DBLP:conf/mascots/JamshidiC16,DBLP:conf/mascots/MendesCRG20,DBLP:conf/sigsoft/JamshidiVKS18,DBLP:journals/corr/abs-2106-02716}. We also compared \approach~with its stationary variant and two state-of-the-art planners for SASs.  Nonetheless, we agree that studying additional systems and other types of planners may prove fruitful.

\section{Related Work}
\label{sec:related}

In this section, we discuss the related work in light of the concepts and design of \approachfs.

\subsection{Stationary Optimization for SASs}

In contrast to the dynamic (and pseudo-dynamic) optimization, stationary planners take a simpler strategy: the search is restarted from scratch whenever there is a new environmental change (or indeed a new search run under a fixed frequency)~\cite{Chen2018FEMOSAA,DBLP:conf/sigsoft/ElkhodaryEM10,DBLP:journals/ase/GerasimouCT18,DBLP:conf/icse/Kumar0BB20,DBLP:conf/icpads/KumarBCLB18,DBLP:conf/kbse/LiXCT20,DBLP:journals/tsc/ChenB17, DBLP:conf/icse/ChenB14}. Among other, Chen \textit{et al.}~\cite{Chen2018FEMOSAA} propose FEMOSAA, a framework that relies on stationary optimization for the SAS planning. While FEMOSAA is designed to make adaptation under a fixed frequency throughout the lifetime of SAS, the search and planning does not run continuously --- every time the search restarts with some randomly generated adaptation plans. Elkhodary \textit{et al.}~\cite{DBLP:conf/sigsoft/ElkhodaryEM10} and Gerasimou \textit{et al.}~\cite{DBLP:journals/ase/GerasimouCT18} also follow similar way with an exact/stochastic search, but the planning is conducted only when there is an environmental change or severe violation of performance requirement has been detected --- yet again each time the search starts from scratch.

Clearly, stationary optimization ignores the information that could have been useful for the next planning, which makes the issues of local optima more difficult to address. With \approachd, we seek to better exploit such information to improve the effectiveness and efficiency of SAS planning.

\subsection{Dynamic Optimization for SASs}

Dynamic optimization for SASs has been traditionally referred to the fact that a vanilla search algorithm runs continuously for planning~\cite{DBLP:conf/icac/RamirezKCM09,DBLP:conf/gecco/0001LY18,DBLP:journals/infsof/ChenLY19,DBLP:journals/taas/KinneerGG21}. For example, Ramirez \textit{et al.}~\cite{DBLP:conf/icac/RamirezKCM09} design \textsc{Plato}, a framework that adopts SOGA for SAS planning. They have shown that without specific designs, SOGA can deal with environmental change by detecting the changes in fitness. As such, \textsc{Plato} meets the concept of lifelong optimization. A similar notion that has been used for SASs is seeding --- the planning and search are ``seeded'' by adaptation plans from previous runs of planning~\cite{DBLP:conf/gecco/0001LY18,DBLP:journals/infsof/ChenLY19}. Kinneer \textit{et al.}~\cite{DBLP:journals/taas/KinneerGG21} follows such a scheme using SOGA for searching in the SAS planner. Conceptually, seeding is similar to running the search and planning continuously and thus is referred to as dynamic optimization for SASs in the literature.

Although the search algorithm is dynamic in nature, the above differs from the true definition of dynamic optimization in the literature~\cite{DBLP:journals/swevo/NguyenYB12}. This is because no specific mechanism has been designed to handle the possible landscape change as the search proceeds, but leaving with the hopes that the algorithm would eventually cope with those changes in some ways. Therefore, the above work can be considered as pseudo-dynamic, and hence the information that can be leveraged after environment change is limited. In contrast, \approach~contains the refined MMO, which is used and designed explicitly to address the formulated multi-modal optimization problem by taking the characteristics of SAS planning into account, aiming to better handle the dynamic changes as the planning runs. As we have shown, preserving multiple local optima (without harming the tendency towards the global optimum) can indeed provide more useful information than simply detecting the changing values of performance objective when the environment changes.

\subsection{Control Theoretical Planning}

Apart from the SBSE, control theoretical approaches have also been applied for SAS planning~\cite{DBLP:conf/sigsoft/FilieriHM15,DBLP:conf/sigsoft/MaggioPFH17,DBLP:conf/sigsoft/ShevtsovW16}. Among others, Maggio \textit{et al.}~\cite{DBLP:conf/sigsoft/MaggioPFH17} use
Kalman filter to revise and update the state values of the controller model, where the core is a model predictive control scheme. Shevtsov and Weyns~\cite{DBLP:conf/sigsoft/ShevtsovW16} also adopt a control theoretical approach, but they extend such with the simplex optimization method, which is essentially a kind of search algorithm.

We consider the contributions in this work as being complementary to the control theoretical planners rather than competitive, because fundamentally the purpose of search and optimization is to seek the global optimum. As a result, \approach~can be integrated with a control theoretical planner such that it seeks to find the global optimum of the current system state modeled by the controller, similar to what has been done by Shevtsov and Weyns~\cite{DBLP:conf/sigsoft/ShevtsovW16}.

\subsection{Performance Learning for SASs}

Another relevant thread of research is performance learning for configurable systems. Various methods have been used, such as neural network~\cite{DBLP:conf/icse/ChenB13,DBLP:conf/icse/HaZ19}, linear regression~\cite{DBLP:conf/icse/ChenB13}, ensemble learning~\cite{DBLP:journals/tse/ChenB17} and transfer learning~\cite{DBLP:conf/kbse/JamshidiSVKPA17}. However, the aim is to learn an accurate function that captures the correlation between adaptation options and the performance while we target planning, therefore our contributions are orthogonal to the above. For example, with an accurate performance model, the Cyber-Twin in \approach~can be replaced by such a model, achieving faster measurements and hence boosting the planning.

\section{Conclusion}
\label{sec:con}

This paper presents \approachd, a lifelong and truly dynamic optimization framework with a hierarchical feedback loop architecture for SAS planning. We formulate the lifelong dynamic optimization for SAS as a multi-modal optimization problem, which is solved by the refined meta multi-objectivization (MMO) model. The aim is to preserve as many local optima as possible without harming the tendency towards the global optimum, hence providing useful information for planning after the environmental change. 

Experiments on three diverse real-world SASs and different environmental changes show that compared with the stationary counterpart and other state-of-the-art planners, \approach~is:

\begin{itemize}
    \item \textbf{more effective}, as it achieves considerably better adaptation plan than the stationary counterparts and other state-of-the-art planners with up to $10\times$ improvement;
    \item and \textbf{more efficient}, since it exhibits $1.4\times$ to $10\times$ speedup on producing promising plans.
\end{itemize}

This work is among the first attempt to explicitly handle the ``dynamic'' on landscape changes during the run of the search for SAS planning.  We show that such a lifelong dynamic optimization is factual for planning in SAS, which can excite a few future research directions, such as considering information from all previous planning runs and memory mechanisms to more precisely track the details of landscape changes between different environments. 

\balance
\bibliographystyle{IEEEtran}
\bibliography{reference} 

\end{document}